# Monitoring cementless femoral stem insertion by impact analyses: an in vitro study

Antoine Tijou[1], Giuseppe Rosi[1], Romain Vayron[1], Hugues Albini Lomami[1], Philippe Hernigou[2], Charles-Henri Flouzat-Lachaniette[2], Guillaume Haïat[1]

[1]CNRS, Laboratoire de Modélisation et de Simulation Multi-Echelle, UMR CNRS 8208, 61 Avenue du Général de Gaulle, Créteil 94010, France

[2]Service de Chirurgie Orthopédique et Traumatologique, Hôpital Henri Mondor AP-HP, CHU Paris 12, Université Paris-Est, 51 avenue du Maréchal de Lattre de Tassigny, 94000 Créteil, France.

**Abstract**

The primary stability of the femoral stem (FS) implant determines the surgical success of cementless hip arthroplasty. During the insertion, a compromise must be found for the number and energy of impacts that should be sufficiently large to obtain an adapted primary stability of the FS and not too high to decrease fracture risk. The aim of this study is to determine whether a hammer instrumented with a force sensor can be used to monitor the insertion of FS.

Cementless FS of different sizes were impacted in four artificial femurs with an instrumented hammer, leading to 72 configurations. The impact number when the surgeon empirically felt that the FS was fully inserted was noted $N_{surg}$. The insertion depth $E$ was assessed using video motion tracking and the impact number $N_{vid}$ corresponding to the end of the insertion was estimated. For each impact, two indicators noted $I$ and $D$ were determined based on the analysis of the variation of the force as a function of time.

The pull-out force $F$ was significantly correlated with the indicator $I$ ($R^2 =0.67$). The variation of $D$ was analyzed using a threshold to determine an impact number $N_d$, which is shown to be closely related to $N_{surg}$ and $N_{vid}$, with an average difference of around 0.2. This approach allows to determine i) the moment when the surgeon should stop the impaction procedure in order to obtain an optimal insertion of the FS and ii) the FS implant primary stability. This study paves the way towards the development of a decision support system to assist the surgeon in hip arthroplasty.

# 1. Introduction

Cementless femoral stem (FS) implants are now widely used in the clinic to restore the functionality of the hip joint. The FS insertion consist in impacting this implant using an hammer in the femur that has been previously reamed with a slightly lower diameter compared to that of the stem, so that it stays in place due to the pre-stressed state of the bone-implant system, a process referred to as "press-fit". Achieving an optimal primary stability of the FS is crucial for the short and long term surgical outcome [1] because micromotions at the bone-implant interface are known to cause early loosening of implants [1] and to jeopardize osseointegration phenomena [2]. In particular, a compromise must be found by the orthopedic surgeon regarding the number and the energy of impacts that should be sufficiently large to obtain a good primary stability of the stem, but that should not be too high to avoid risks of intraoperative (with a reported incidence of 4.1-27.8% [3-7]) and post-operative [8-10] peri-prosthetic femoral fracture. So far, the orthopedic surgeons use empirical methods such as their proprioception to estimate the optimal number and energy of impacts in order to reach the aforementioned compromise. In particular, surgeons listen to the acoustic signature of the impact between the hammer and the ancillary [11] to adapt their surgical strategy. It remains difficult to quantitatively estimate the primary stability of the FS in the clinic.

To the best of our knowledge, three methods have been employed in the literature in order to assess the primary stability of the FS [1]. First, micromotions at the bone-implant interface were measured using linear variable differential transducers (LVDT's), which was used to compare the FS primary stability for three kinds of stem fixation [12] and for resurfacing hip implants under different loading conditions [13] in composite bone (mimicking the human bone mechanical properties). LVDt's were also used in order to compare the biomechanical behavior of FS with different sizes or shapes while inserted in fresh human femur [14, 15] and in composite bone [16]. However, the protocol using LVDt's cannot be used in the operative room.

Second, micro-computer tomography (µCT) imaging has been used to quantify micromotions at the bone-implant interface through a method developed by Gortchacow et al. [17, 18] to analyze the effect of a collar on the FS primary stability [19] in cadaveric femurs. However, errors related to metal artefacts have been noticed, which may alter the clinical use of this approach in the operating room.

Third, a method based on vibrational analysis of the FS [20] has been developed by Pastrav et al. in order to follow the FS insertion *in vivo* [21] and in composite bone [22]. The main difference between the approach developed by Pastrav et al. [20-22] and the one presented herein is that the present study considers the use of a dedicated impact hammer that is used to carry insert the femoral stem inside the femur.

In previous studies by our group, an approach based on the analyses of the impacts between the hammer and the ancillary has been developed by our group in order to assess the acetabular cup (AC) primary stability. First studies based on mass drops [23-25] and using bovine bone samples showed that the contact duration was correlated to the AC implant insertion. Then, an indicator based on the impact momentum was shown to be more accurate to assess the implant primary stability [25]. Based on these results, an instrumented hammer using a dynamic force sensor screwed on the impacting face of the hammer was developed, and the technique was adapted to predict the AC implant stability [26]. Moreover, static [27, 28] and dynamic [29] finite element models have been developed to understand the mechanical phenomena occurring during and after the AC implant insertion. Eventually, a cadaveric study showed that the instrumented impact hammer could be used in situations closed to the clinics in order to determine the AC implant primary stability [30]. However, using an impact hammer to investigate the biomechanical behavior of the FS has not yet been investigated.

The aim of this study is to determine whether an instrumented impact hammer can be used to monitor the insertion of cementless FS. The originality of the method described herein compared to the other approaches described in the literature such as vibrational analysis, microcomputed tomography imaging or LVDT micromotion measurements lies in the use of an instrumented hammer which is used similarly as the surgical hammer employed in the operating room so far. In particular, we aim at estimating the variation of the signal retrieved during impacts as a function of the impact number. Another objective is to determine whether it is possible to use this impact hammer to determine the optimal position of the FS in the femur and to assess the FS biomechanical stability. To do so, six composite artificial femurs were used and the instrumented hammer was employed throughout the insertion procedure. An optical system was used to follow the insertion of the FS using video motion tracking techniques. The orthopedic surgeon was asked to determine using his proprioception when he empirically felt that the stem was fully inserted. Then, different impacts were realized and the impact momentum was determined and compared with the FS pull-out force.

## 2. Materials and methods

2.1. Specimens and implants

We used cementless FS (CERAFIT R-MIS) manufactured by Ceraver (Roissy, France). Five sizes were used (from size 7 to 11). The optimal size is equal to 9 according to the surgeon but different sizes were considered in order to consider different stability conditions. The FS are made of titanium alloy (TiAl6V4) with a hydroxyapatite coating – except on the neck – in order to facilitate osseointegration phenomena. The corresponding reamers manufactured by Ceraver (Roissy, France) were used prior to the FS insertion. During the insertion of the FS, the ancillary was screwed directly within the FS in order to obtain a rigid bilateral fixation between the FS and the ancillary, as shown in Fig. 1.

Six identical artificial femurs named "OrthoBones" (3B Scientific, Hamburg, Germany) were employed herein. These sample are made of two composite materials (polyurethane foam) that mimic the biomechanical properties of human trabecular and cortical bone tissue. The samples were prepared as follow (see Fig. 1). First, each sample was cut at the diaphysis level, in the middle of femur. Then, the distal end of the sample was embedded in a fast-hardening resin (SmoothCast 300 polymer, Smooth-On, Easton, PA, USA). Finally, the osteotomy of the femoral neck was performed following the usual clinical protocol. All experiments were carried out by an experienced orthopedic surgeon.

2.2. Instrumented hammer

The hammer (m = 1.3 kg) used to insert the FS was the same as the one used in our previous studies dealing with the acetabular cup insertion [26, 30]. A dynamic piezoelectric force sensor (208C05, PCB Piezoelectronics, Depew, New York, USA) was screwed to the center of the impacting face of a carpenter's hammer, similarly as previous studies by our group dealing with the acetabular cup implant [24-26, 30, 31]. All impacts were realized directly on the force sensor so that the variation $s(t)$ of the force as a function of time could be measured. The measurement range goes up to 22 kN in compression.

2.3. Experimental set-up

2.3.1. Femoral stem insertion

Figure 1 describes schematically the experimental setup. In order to find an experimental configuration that mimics the clinical situation as accurately as possible, a 1 mm thick of 200 bloom porcine gelatin layer (Luis Francois, Croissy Beaubourg, France) was placed underneath the resin block described in subsection 2.1. The cylindrical resin part was embedded in textile tissue that was positioned in a vise in order to avoid any movement perpendicularly to the femur axis. Following the surgeon proprioception, this configuration allows to obtain an acoustical and mechanical behavior of the FS qualitatively comparable to what is obtained clinically. In particular, this system has a vertical degree of freedom (due to the presence of gel), which is necessary to avoid obtaining a rigid fixation that may significantly modify the impact conditions.

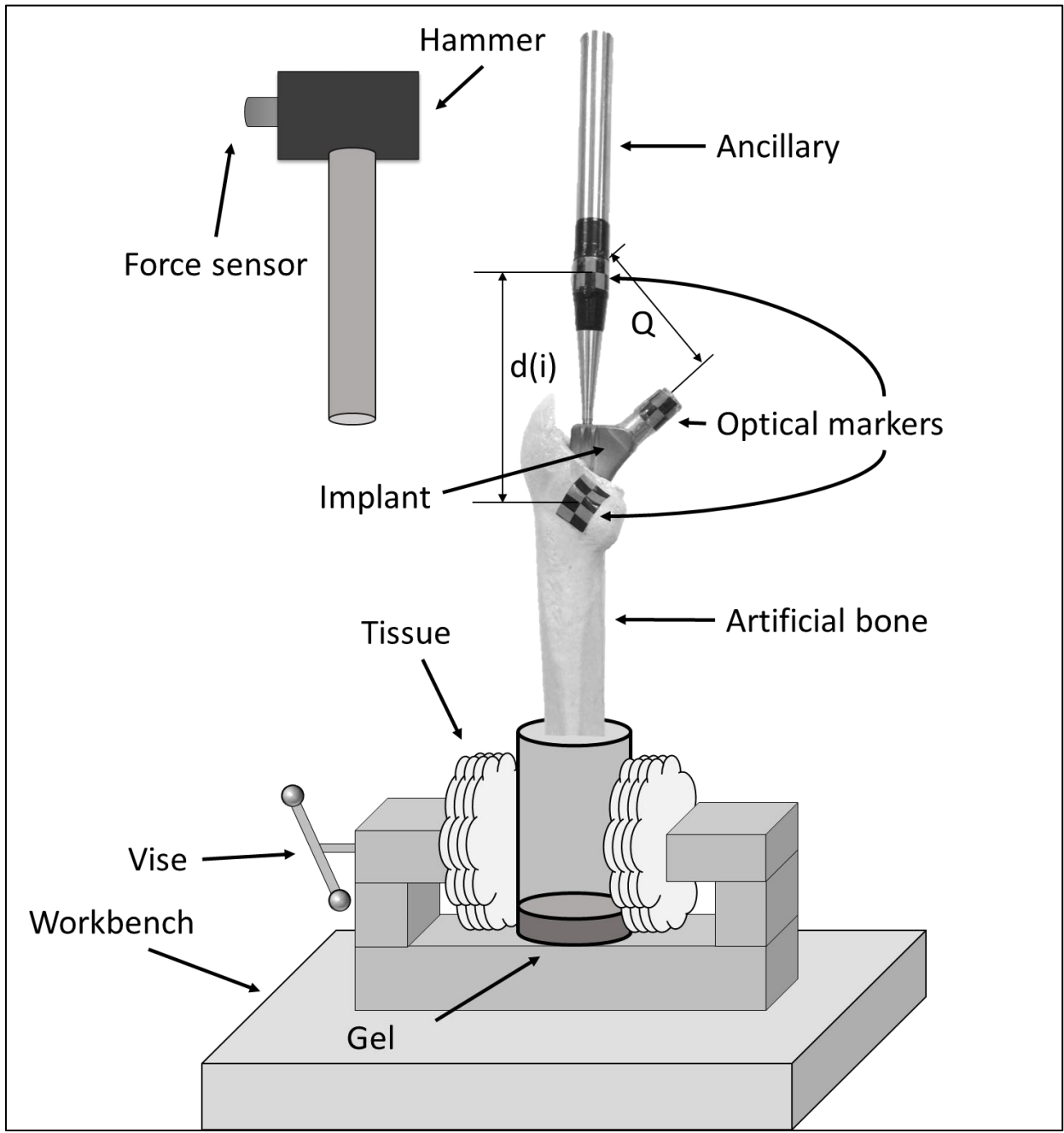

*Figure 1. Schematic representation of the experimental configuration.*

Each impact was realized using the impact hammer described in subsection 2.2 and a data acquisition module (NI9234, National Instruments, Austin, TX, USA) with a sampling rate of 51.2 kHz and a resolution of 24 bits was used to record the time-variation of the force, denoted s(t), applied between the hammer and the ancillary. Then the data were transferred to a computer and recorded using a Labview interface (National instruments, Austin, TX, USA) during a duration of 2 ms.

2.3.2. Video motion tracking

In order to follow the insertion of the FS, a camera (Powershot SX410 IS, Canon, Tokyo, Japan) with a frame rate of 24 frames per second and a high-definition resolution of 1280x7200 MP was attached to a tripod and filmed the bone-implant system throughout the impaction procedure. Optical markers were placed on the FS implant (on the femoral neck), on the ancillary and on the artificial bone (see Fig. 1). The relative displacements of the markers after each impact was determined with the software Tracker (Cabrillo College, Aptos, CA, USA).

The indicator $E$ corresponding to the relative movement of the FS compared to the artificial bone was obtained by analyzing the video and was defined as:

$$E(i) = d(0) - d(i) \qquad (1)$$

where $d(i)$ corresponds to the distance between the markers located on the artificial bone and on the ancillary after the $i^{th}$ impact (see Fig. 1) The constant distance $Q$ defined in Fig. 1 was used to convert pixel into centimeters.

2.3.3. Femoral stem extraction

The pull-out force $F$ was determined by the maximum value of force obtained when pulling out the FS from the host artificial bone in the direction of the axis of the stem using a mechanical testing machine (DY25, Adamel Lhomargy, Roissy en Brie, France). The rigid resin block was fixed and a constant displacement velocity (3.3e-5 m/s) was applied to the ancillary using the testing machine crosshead that was attached to its end. Note that the other end of the ancillary was screwed into the FS proximal end.

2.4. Signal processing

A dedicated signal processing technique was developed in order to extract information from the signal $s(t)$ defined in subsections 2.2. and 2.3. The beginning of the impact ($t$=0) was defined when the signal first exceeded a threshold of 200 N. Two different signal processing methods were considered for each signal corresponding to each impact.

First, the time difference (noted $D$ in what follows) between the time of the second and of the first local maxima of $s(t)$ was determined following:

$$D = max2\big(s(t)\big) - \max(s(t)) \qquad (2)$$

where the function *max* denotes the time of the maximum value of the signal, named first peak, and the function *max2* denotes the time of the second local maximum, named second peak, for which the following conditions are fulfilled, which were applied in order to avoid considering second peak that may correspond to false alarm situations. First, the prominence of the considered peak (denoted $\alpha$), corresponding to the amplitude difference between the peak and the closest local minimum, must be higher than 100 N. Second, the time difference between the first peak (defined above) and the considered peak (denoted $\beta$), must be higher than 0.1 ms. Third, the parameter $\gamma$, representing the width of the peak at the half of the prominence $\alpha$, must be higher than 0.04 ms. The choice of the parameters $\alpha$, $\beta$ and $\gamma$ will be discussed in section 4.

Second, the indicator $I$ was defined as the impact momentum similarly as in [25, 26, 30] following:

$$I = \frac{1}{(t_2 - t_1)} \int_{t_1}^{t_2} s(t)dt, \qquad (3)$$

where $t_1$ = 0.19 ms and $t_2$ = 0.31 ms define the time window of integration. Again, the choice of the values of $t_1$ and $t_2$ will be discussed in section 4.

2.5. Experimental protocol

Figure 2 summarizes the experimental protocol that was carried out by a trained orthopedic surgeon for each artificial bone sample and that is described in more details in what follows.

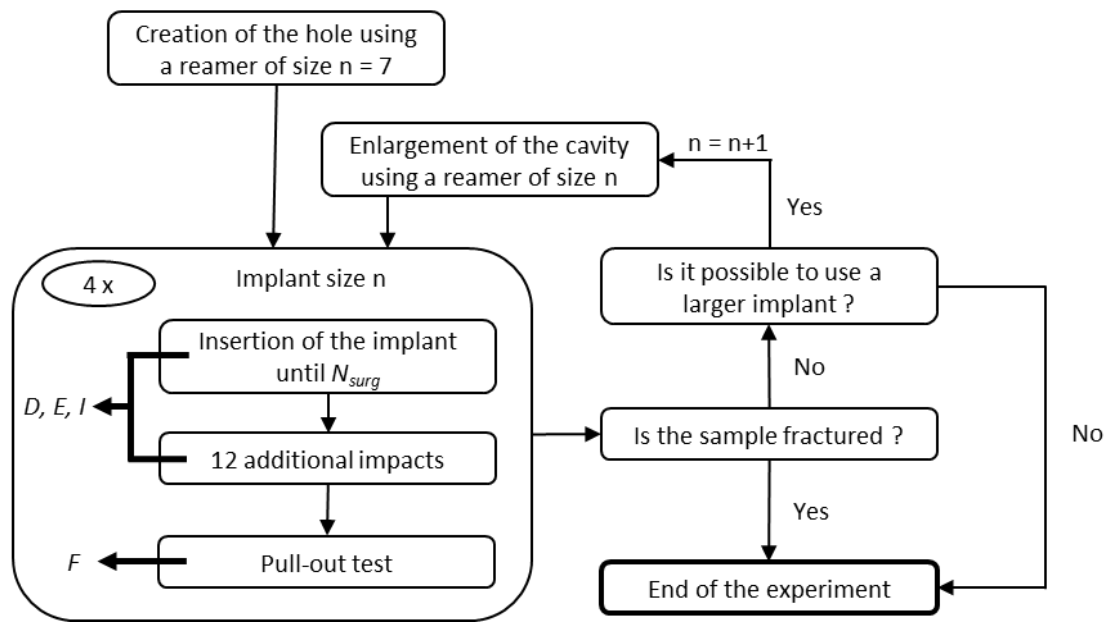

*Figure 2. Schematic representation of the protocol realized for each artificial bone sample.*

A cavity was initially created in each sample using the reamer with the smallest size available (size n = 7). Then, the following impaction procedure was carried out, which consists in three consecutive steps. Firstly, the surgeon inserted the FS into the sample using the instrumented impact hammer until he empirically considered that an optimal stability condition was obtained. The number of impacts needed to obtain this "optimal" stability condition was noted $N_{surg}$. Secondly, twelve additional impacts with an amplitude of the first varying between 2 and 9 kN were performed in order to determine i) whether the FS could be further inserted within the sample and ii) the value of the indicator *I* after the insertion. The choice of the values used for the first peak amplitude and for the number of impacts will be discussed in section 4. For each impact, the values of the indicators *D*, *I* and *E* were determined. Thirdly, the implant was extracted axially and the value of the pull-out force *F* was determined.

This same impaction procedure was repeated four times. Then, the surgeon checked whether the sample was fractured. If so, the experiment was ended for this sample. Otherwise, a larger implant was used if the surgeon judged that inserting a larger implant was possible. Ifnot, the experiment ended. In all cases, the corresponding reamer with the same size than the FS was used at all times.

2.6. Post-processing and method of data analysis

Following the experimental protocol described above, three different methods were employed for each insertion procedure in order to estimate the number of impact necessary to "fully" insert the FS within the femur. This "full" insertion thus corresponds to an estimation of the end of the migration phase of the FS into the femur.

The first method consists in analyzing the variation of the parameter *D(i)* corresponding to the time difference between the first and second maxima as a function of the impact number *i*. For each insertion procedure, the parameter $N_d$ was defined as the number of the first impact that satisfies the following inequality:

$$D(i) \leq D_{th} \qquad (4)$$

where $D_{th}$ is a threshold chosen equal to 0.53 ms, which will be discussed in section 4.

The second method consists in analyzing the variation of the parameter *E(i)* corresponding to the position of the FS relatively to the femur as a function of the impact number *i*. For each insertion procedure, the parameter $N_{vid}$ was defined as the number of the first impact that satisfies the following inequality:

$$E(i) \geq E_m - \delta * E_{sd} \qquad (5)$$

where $E_m$ and $E_{sd}$ are respectively the average and standard deviation of the values of *E* for the last eleven impacts and δ is a parameter empirically chosen equal to 3.5. The value of the parameter δ will be discussed in section 4.

The third method was described in subsection 2.5 and consists in using the proprioception (touch, sight and hearing) of the surgeon, similarly as what is done in the clinic, to determine when he felt that the FS was totally inserted, leading to the parameter $N_{surg}$.

The differences between the values found for $N_{surg}$, $N_d$ and $N_{vid}$ were determined for each insertion procedure, leading to:

$$M_d = N_d - N_{surg}, \quad M_{vid} = N_{vid} - N_{surg}, \quad M_c = N_{vid} - N_d \tag{6}$$

Then, for each insertion procedure, the average and standard deviation values of the indicator $I$ (denoted $I_m$ and $I_{sd}$) were determined. In order to obtain comparable results, only the four signals with the first peak amplitude closest to 4700 kN were selected. Choices of the number of signals and of the central first peak amplitude will be discussed in section 4. A linear regression was applied between $I_m$ and $F$. Note that each composite femur was used many times (as shown in Table 1), which may lead to the presence of microcracks, this point being discussed in section 4.

3. **Results**

Table 1 shows the number of impaction procedures considered for each artificial bone sample.

Figure 3 shows seven examples of force signals s(t) obtained for a given configuration (sample #1, size of implant = 9 and test #4). The respective number of each impact is indicated above each corresponding second peak of the signal. The signals shown correspond to the second, third, fourth, fifth, ninth and sixteenth impact. As illustrated in Fig. 3, the time of the second peak first decreases and then stays constant after the ninth impact.

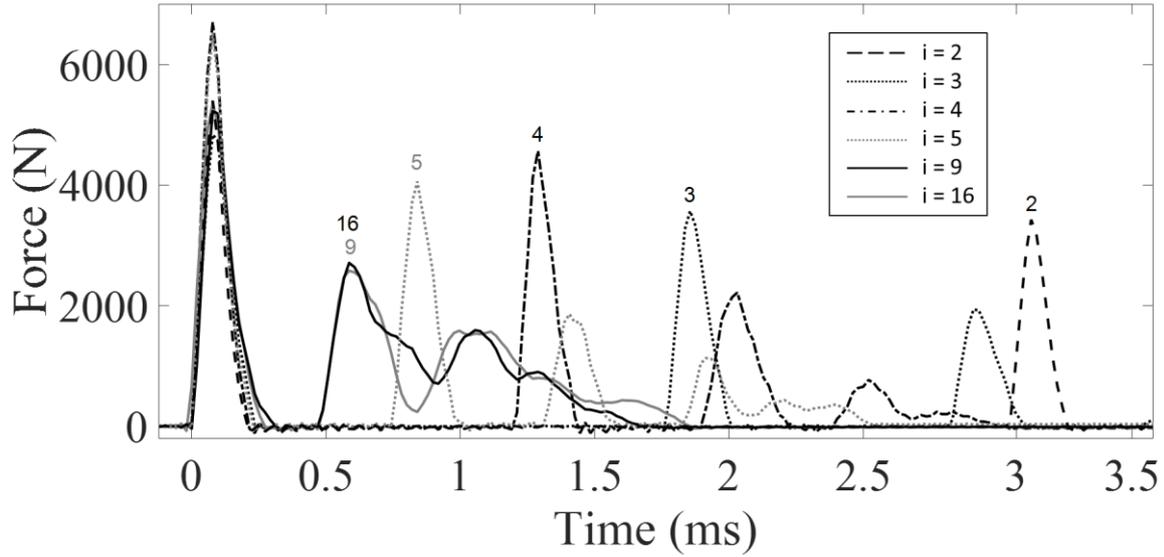

*Figure 3. Seven signals corresponding to the time variation of the force obtained during the impact of the instrumented hammer on the ancillary linked to the femoral stem (sample #1, size of implant 9, test #4).*

| Size of implant | Sample #1 | Sample #2 | Sample #3 | Sample #4 | Sample #5 | Sample #6 | Total |
|---|---|---|---|---|---|---|---|
| 7 | 4 | 4 | 4 | 4 | 4 | 4 | 24 |
| 8 | 4 | 4 | 4 | 2 | 4 | 4 | 22 |
| 9 | 4 | 4 | 4 | 0 | 4 | 4 | 20 |
| 10 | 2 | 0 | 0 | 0 | 0 | 4 | 6 |
| Total | 14 | 12 | 12 | 6 | 12 | 16 | 72 |

*Table 1. Number of configurations performed in this study for each size of implant for each samples, for a total of 72 configurations considered in this study*

Figure 4 shows the variation of the parameters $D$ and $E$ as a function of the impact number $i$ for the same configuration as the one corresponding to Fig. 3. The vertical dashed black line represents the reference value given by the surgeon, namely $N_{surg} = 9$, which corresponds to the insertion endpoint determined by the surgeon for the specific configuration shown in Fig. 4. The horizontal dashed (respectively dashed) black line represents the penetration equal to $E_m - \alpha \times E_{sd}$ (respectively the threshold $D_{th} = 0.53$ms). For this configuration $N_{surg} = N_d = N_{vid}$, so $M_d = M_{vid} = M_c = 0$.

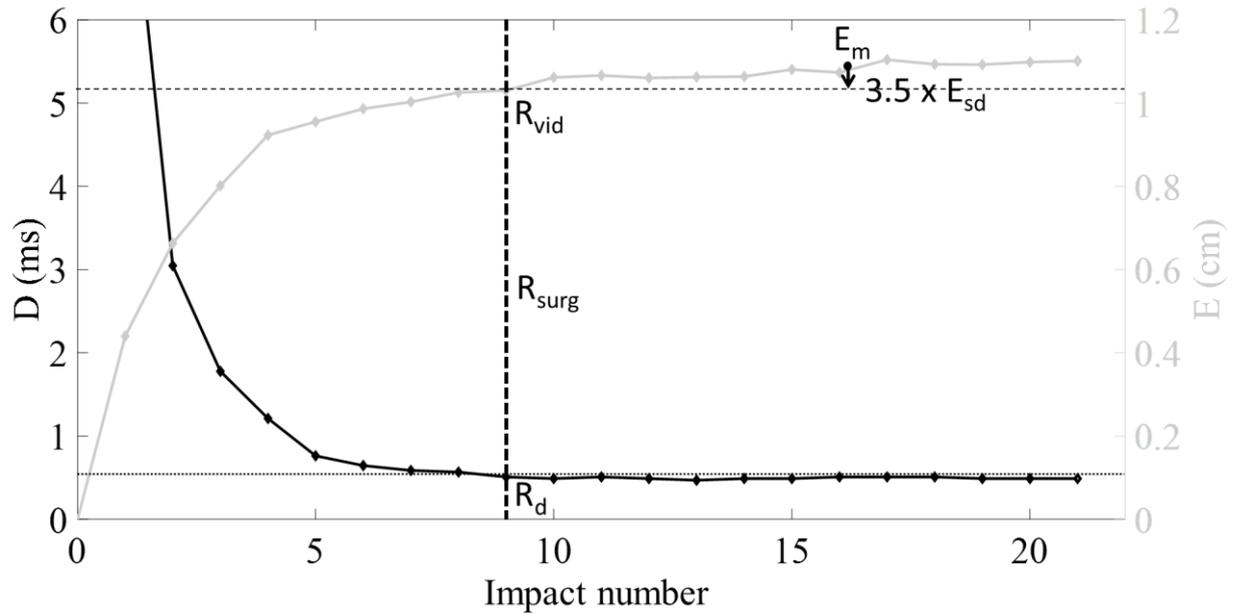

*Figure 4. Variation of parameters D (black) and E (grey) corresponding respectively to the difference of the time of the first and second maxima and to the implant penetration depth as a function of the impact number for the same configuration as the one corresponding to Fig. 3. The vertical dashed black line represents $N_{surg}$, the horizontal dashed black line represents the penetration equal to $E_m - α \times E_{sd}$ and the horizontal dotted black line represents the threshold $D_{th} = 0.53ms$. For this configuration, $N_{surg} = N_{vid} = R_d = 9$ was obtained.*

Figure 5 shows the distributions of the results obtained for $M_d$, $M_{vid}$ and $M_c$, which corresponds to the difference obtained between the three different methods for the estimation of the insertion endpoint of the FS in the sample. Table 2 shows the average and standard deviation values obtained for $M_d$, $M_{vid}$ and $M_c$, which are in the same range of variation.

|      | Average | Standard deviation |
|------|---------|--------------------|
| Md   | -0.26   | 1.16               |
| Mvid | -0.43   | 1.51               |
| Mc   | -0,17   | 1.54               |

*Table 2. Average and standard deviation values of $M_d$, $M_{vid}$ and $M_c$.*

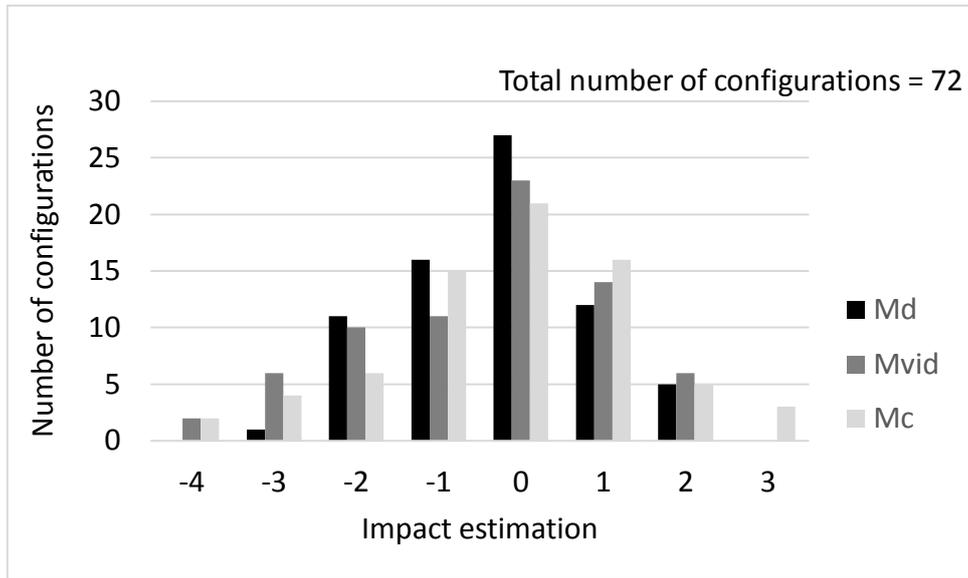

*Figure 5. Distribution of the values obtained for $M_d$, $M_{vid}$ and $M_c$*

Figure 6 shows four signals corresponding to four different configurations. Each signal shown in Fig. 6 corresponds to the averaged value of the four signals $s(t)$ after $N_{surg}$ with a first peak amplitude closest to 4.7 kN. The vertical black lines correspond to times $t_1$ and $t_2$, which were used as bounds of the time window employed for the definition of the impact momentum. The dashed vertical gray line on the left (respectively right) represents the time of the first peak (respectively the threshold $D_{th} = 0.53$ ms). The corresponding values of the pull-out force $F$ and of the indicator $I_m$ are indicated for each configuration.

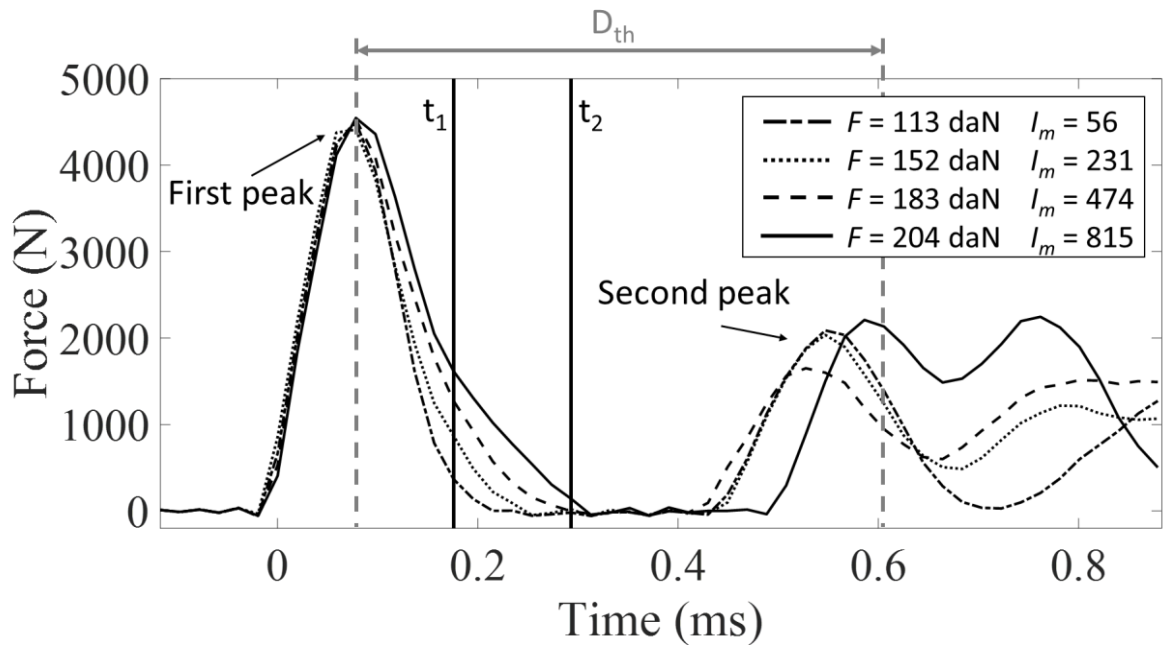

*Figure 6. Four averaged signals obtained when the femoral stem is fully inserted in the artificial bone sample corresponding to four configurations. The corresponding values of F and of $I_m$ are indicated.*

The relationship between the indicator $I_m$ and the pull-out force $F$ is shown in Figure 7, together with the results of a linear regression analysis. A significant correlation is obtained between the indicator $I_m$ and the pull-out force and the determination coefficient ($R^2$) is equal to 0.67. The errorbars correspond to the standard deviations $I_{sd}$ obtained for each configuration.

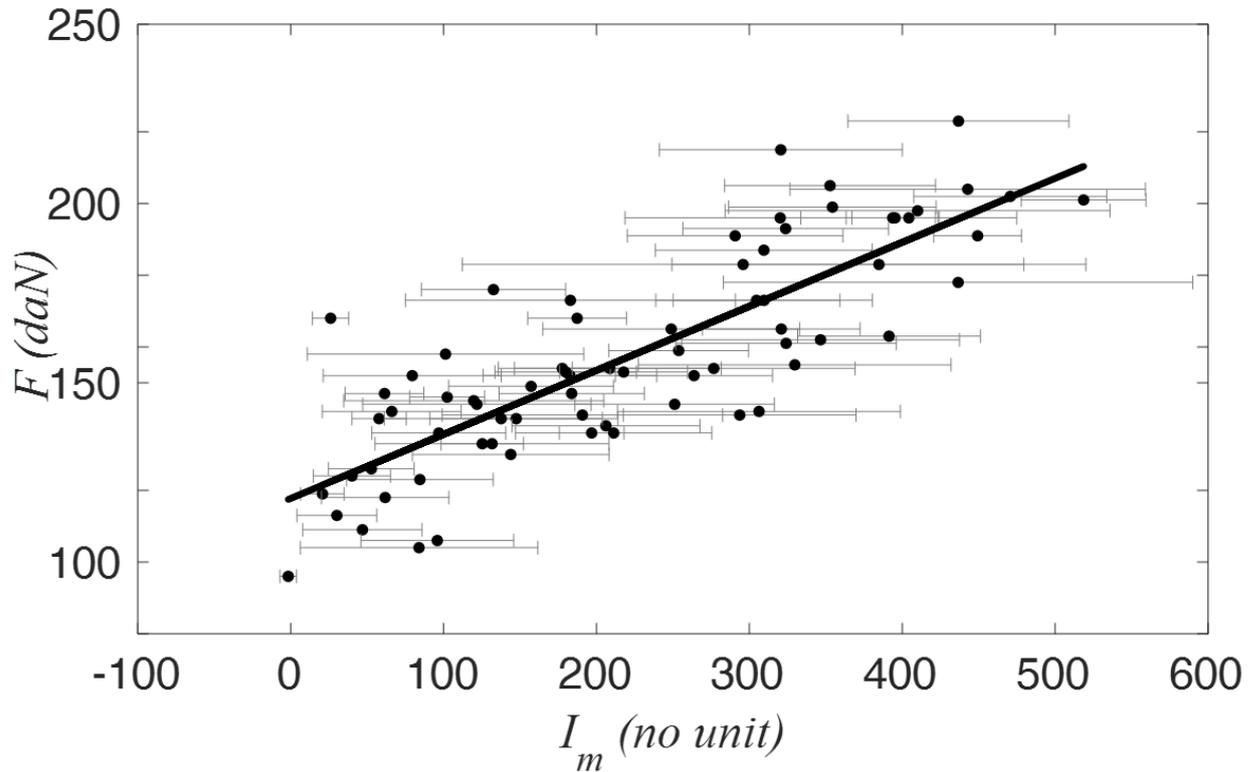

*Figure 7. Relationship between the averaged value of the indicator $I_m$ and the pull-out force F. The determination coefficient is $R^2 = 0.67$. The horizontal error bars correspond to the values of the standard deviation of the indicator $I_{sd}$.*

4. **Discussion**

The main originality of the present study is to use an instrumented hammer to follow the femoral stem insertion and assess its insertion endpoint in an artificial bone sample. Another originality is to quantify the difference between the insertion endpoint estimated using i) the impact hammer, ii) video motion tracking and iii) the surgeon proprioception. Note that this approach is comparable to what has been done for the acetabular cup in [25, 26, 30], except that the signal processing technique as well as the aim (determination of the insertion endpoint) differ when considering the FS and the AC implant. However, the same hammer was used for the AC and FS implant insertion.

Figures 3 and 4 shows that the value of the indicator $D$, which corresponds to the time of the second peak of the signal decreases as a function of the number of impacts. When the FS is inserted into the host bone, the bone-implant contact ratio increases, which leads to an increase of the overall rigidity of the bone-implant system. The increase of the rigidity of the system may in turn explain the increase of its resonance frequency, which leads to a decrease of the difference between the first and second peak of the signal.

Figure 7 shows that the indicator $I_m$ and the pull-out force $F$ are significantly correlated, which indicates that the analysis of the time-variation of the force during the impact allows to quantitatively assess the primary stability of the femoral stem. This results ($R^2 = 0.67$) is qualitatively similar to the results obtained for the acetabular cup using the same impact hammer in an *in vitro* study ($R^2 = 0.83$) [26] and in a cadaveric study ($R^2 = 0.69$) [30].

One of the aim of the present study was to determine the insertion endpoint of the FS, which corresponds to the moment during the impaction procedure when the FS is fully inserted into the artificial bone sample. To do so, the results obtained with the impact signal analyses were compared with two more 'classical' methods. First, the proprioception of the experienced surgeon who carried out the experiments was considered (leading to the estimate $N_{surg}$), similarly as what is done in the clinic. Although informative, this method has the drawback of depending on the surgeon and to be associated to possible bias, which were estimated empirically by the surgeon equal to around +-2 impacts approximately. Second, the method based on video motion tracking also suffers from errors, which are due i) to changes of angular position, ii) possible macroscopic 3D movements and iii) errors based on the image processing technique. Moreover, as shown in Fig. 4, the value of $E$ continues to slightly increase, even after the insertion endpoint is deemed to be reached by the surgeon. The increase of $E$ could be related to possible plastic deformations of the host bone, which may occur even after the insertion endpoint is reached. The continuous increase of $E$ explain the choice of the parameters indicated in Eq. 5. Therefore, the aforementioned factors lead to an error on $N_d$ equal to around +-2. The errors related to the two aforementioned techniques may explain the results shown in Fig. 5, which shows that the difference obtained between the three techniques is of the order to magnitude of the cumulative errors described above. Despite the presence of the aforementioned errors, a relatively good agreement is obtained between the three different methods (see Fig. 5), which constitutes a validation of the approach.

In this study, several parameters were chosen empirically. First, the choice of the number of additional impacts given by the surgeon (equal to twelve) was the result of a compromise between i) a sufficiently high number to be certain to obtain enough signals to assess the reproducibility of the indicator $I$ and to obtain a convergence for the variation of the indicators $D$ and $E$ and ii) a sufficiently low number to minimize fracture risk. The upper bound of the range of variation [2 – 9 kN] of the maximal force of the twelve impacts realized once the surgeon felt that the FS is fully inserted (i.e. after $N_{surg}$) was chosen sufficiently low in order to minimize fracture risk. This range of variation (7 kN) was chosen sufficiently wide in order to be able to determine the influence of the maximal force on the value of $D$. Note that this range of variation is similar to the typical range applied by the surgeon during the insertion, i.e. before $N_{surg}$. Figure 8 shows the variation of the values of $D$ obtained for the 864 (corresponding to 12*72) impacts realized after $N_{surg}$. The dotted line corresponds to the threshold equal to 0.53 ms chosen for $D_{th}$. As shown in Fig. 8, no significant variation of the value of $D$ was obtained as a function of the first peak amplitude. Moreover, the value of $D$ was higher than $D_{th}$ for only 10 impacts out of 864, which constitutes a validation of the approach and explains the choice of $D_{th}$.

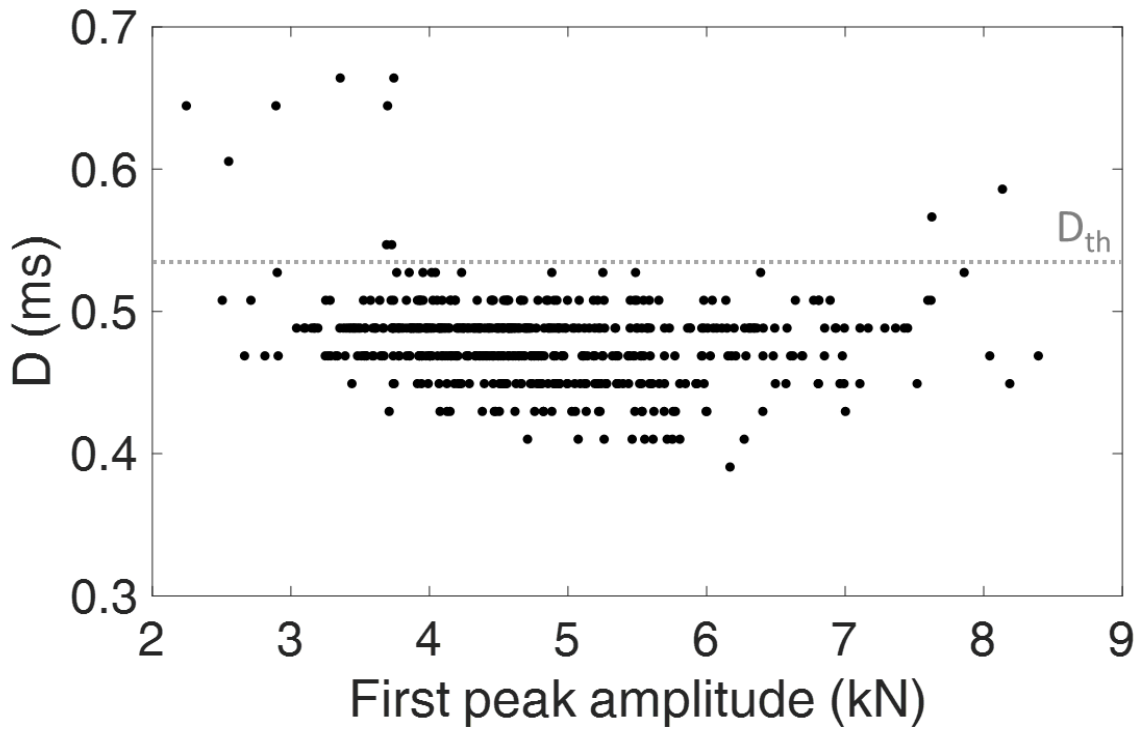

*Figure 8. Variation of the time D of the second peak of the signal for all impacts realized after $N_{surg}$ (i.e. when the femoral stem is fully inserted) for all configurations as a function of the amplitude of the first peak. The horizontal dotted grey line represents the threshold $D_{th}$ = 0.53ms chosen to determine when the femoral stem is fully inserted ($N_d$).*

Second, the values of $t_1$ and $t_2$ used to compute the indicator $I$ have been obtained following an optimization study carried out to determine the values of $t_1$ and $t_2$ that maximize the determination coefficient of the linear regression between $I_m$ and $F$. Varying the value of $t_1$ between 0.17 ms and 0.21 ms or the value of $t_2$ between 0.25 ms and 0.45 ms did not alter significantly the results (less than 2% difference for $R^2$, data not shown). Moreover, the number of impacts considered to determine $I_m$ (equal to four) was chosen sufficiently high to obtain a consistent value of $I_m$. The selected impacts corresponded to the impacts having a first peak amplitude closest the overall median value equal to 4.7 kN. Despite this choice, relatively high standard deviation values $I_{sd}$ were found (see errorbars in Fig. 6), which may be explained by the relatively large range of variation of the values of the first peak amplitude [3976 N – 5361 N]. Reducing this range of variation should decrease the values of $I_{sd}$, which should be checked in a future study.

Third, the impacts used to compute $E_m$ were the last eleven impacts after $N_{surg}$. Moreover, the parameter $\delta$ (equal to 3.5) was chosen empirically to find a compromise between a sufficiently high value to account for the constant increase of $E$ after $N_{surg}$ and a sufficiently low value in order not to underestimate $E_m$. Varying the value of $\delta$ between 2.8 to 4.5 did not alter the value of $M_{vid}$ (less than 10%, data not shown).

Fourth, the value of 200 N used for the threshold defining the beginning of the contact between the hammer and the ancillary was chosen to find a compromise between a sufficient value compared to the signal to noise ratio and a value small enough to provide a good accuracy of detection. Changing the value of the detection this threshold between 150 and 250 N did not modify the results.

Fifth, the parameters used to detect the second peak were chosen in order to avoid any false alarm corresponding to erroneous position of the peak. Variations of $\alpha$ between 50 N and 150 N, $\beta$ between 0.02 ms and 0.06 ms and $\gamma$ between 0.06 ms and 0.14 ms do not change the value of $N_d$.

The main limitation of this study lies in the use of bone mimicking phantoms instead of actual human femurs, which is likely to modify the mechanical response of the bone-implant system because of the difference in terms of bone properties as well as because of the absence of the surrounding soft tissues. Note that in the case of the AC implant, we verified that changing the soft tissue thickness did not change significantly the results [31], which should be checked for the present configuration. Moreover, all phantoms were similar and did not exhibit any variability. Microcracks may appear in the composite femurs and it may be difficult to detect them, which corresponds to a limitation of our approach. However, we verified visually that no apparent crack could be detected after each configuration. Note that other studies [16, 32, 33] have already used composite femurs in comparable situations. Future step will consist in using cadaveric specimen in order to determine the performances of the method in conditions closer to the operating room. A particular attention will be brought to the possibility to determine one given value of $D_{th}$ for all samples.

5. **Conclusion**

This study shows that the analysis of the time variation of the force between the hammer and the ancillary for each impact during the insertion of the femoral stem allows to assess the number of impacts necessary to obtain an adapted insertion of the implant in the host bone. This method, associated with an adapted instrumented hammer could constitute the principle of a future medical device consisting in a real time decision support system helping surgeons to adapt their surgical strategy in a patient specific manner. More specifically, an alert could be implemented and activated when the femoral is expected to be fully inserted and stable, thus avoiding unnecessary additional impacts which may lead to fracture risks. A significant advantage of the current method lies in that it can be used with minimal changes of the surgical protocol.

**Acknowledgement**

This work has received funding from the European Research Council (ERC) under the European Union's Horizon 2020 research and innovation program (grant agreement No 682001, project ERC Consolidator Grant 2015 BoneImplant).

This work has received funding from the CNRS through the PEPS INGENIERIE TRANSLATIONELLE EN SANTE 2017 project "HipImpact".